\documentclass[lettersize,journal]{IEEEtran}
\IEEEoverridecommandlockouts
\usepackage{cite}
\usepackage{amsmath,amssymb,amsfonts}
\usepackage{algorithmic}
\usepackage{graphicx}
\usepackage{textcomp}
\usepackage{xcolor}
\usepackage{multicol}
\usepackage{siunitx}
\usepackage{tablefootnote}
\usepackage{color}
\newcommand{\revision}[1]{{\color{black}{#1}}}
\newcommand{\revisiontwo}[1]{{\color{black}{#1}}}

\usepackage[normalem]{ulem}

\begin{document}

\onecolumn
\textcopyright~2024 IEEE.  Personal use of this material is permitted.  Permission from IEEE must be obtained for all other uses, in any current or future media, including reprinting/republishing this material for advertising or promotional purposes, creating new collective works, for resale or redistribution to servers or lists, or reuse of any copyrighted component of this work in other works.
\newpage
\twocolumn
\title{}

\title{\revision{A Comparative Analysis of Smartphone and Standard Tools for Touch Perception Assessment Across Multiple Body Sites}


\thanks{Manuscript received XXXXXX XX, XXXX; revised XXXXXX XX, XXXX;
accepted XXXXXX XX, XXXX. Date of publication XXXXXX XX, XXXX; date of
current version XXXXXX XX, XXXX. This article was recommended for publication by Associate Editor XXXXXXXXXXXX and Editor-in-Chief XXXXXXXXXXXX upon evaluation of
the reviewers’ comments.}
\thanks{R.\ A.\ G.\ Adenekan, K.\ T.\ Yoshida, and A.\ M.\ Okamura are with the Mechanical Engineering Department,  A.\ Gonzalez Reyes is with the Computer Science Department, \revision{and S. Kodali is with the Electrical Engineering Department} at Stanford University, Stanford, CA 94305.  (email: \{adenekan; kyle3; aokamura; alegre; \revision{kodali}\}@stanford.edu).} 
\thanks{C.\ M.\ Nunez is with the Sibley School of Mechanical and Aerospace Engineering at Cornell University, Ithaca, NY 14853. (email: cmn97@cornell.edu).}
\thanks{This work was supported in part by a grant from the Precision Health and Integrated Diagnostics Center at Stanford, the National Science Foundation Graduate Fellowship Program, and the Stanford Graduate Fellowship Program.}

\author{Rachel A.\ G.\ Adenekan,~\IEEEmembership{Student Member,~IEEE}, Alejandrina Gonzalez Reyes,\\ Kyle T.\ Yoshida,~\IEEEmembership{Student Member,~IEEE}, \revision{Sreela\ Kodali,~\IEEEmembership{Student Member,~IEEE},}\\ Allison M.\ Okamura,~\IEEEmembership{Fellow,~IEEE,} and Cara M.\ Nunez,~\IEEEmembership{Member,~IEEE}

\vspace{-7pt}

}

}

\maketitle

\begin{abstract}
Tactile perception plays an important role in activities of daily living, and it can be impaired in individuals with certain medical conditions. The most common tools used to assess tactile sensation, the Semmes-Weinstein monofilaments and the 128~Hz tuning fork, have poor repeatability and resolution. Long term, we aim to provide a repeatable, high-resolution testing platform that can be used to assess vibrotactile perception through smartphones without the need for an experimenter to be present to conduct the test. We present a smartphone-based vibration perception measurement platform and compare its performance to measurements from standard monofilament and tuning fork tests. We conducted a user study with \revision{36 
}healthy adults in which we tested each tool on the hand, wrist, and foot, to assess how well our smartphone-based vibration perception thresholds (VPTs) detect known trends obtained from standard tests. The smartphone platform detected statistically significant changes in VPT between the index finger and foot and also between the feet of younger adults and older adults. Our smartphone-based VPT had a \revision{moderate
}  correlation to tuning fork-based VPT. Our overarching objective is to develop an accessible smartphone-based platform that can eventually be used to measure disease progression and regression. 

\end{abstract}

\begin{IEEEkeywords}
smartphone, vibration, tactile, perception
\end{IEEEkeywords}

\section{Introduction}
Tactile perception, including vibrotactile perception, plays a critical role in enabling humans to perform various sensorimotor tasks such as object manipulation, navigation, and playing sports~\cite{alahakone2009real,evreinov2004mobile,koehn2015surgeons,SHULL201411}. We even use vibrotactile perception to balance during walking, a pervasive activity of daily living~\cite{richardson1992relationship,o1993incidence,era1996postural,tiedemann2010development,hafstrom2018perceived}. Many underlying health conditions \revisiontwo{and treatments} including diabetes, chemotherapy, and direct injuries to the body can impair our tactile perception~\cite{maiya2020relationship}. Given our reliance on tactile perception, deficits in perceiving tactile cues can have devastating consequences. Assessing tactile perception may help us understand disease progression and recovery, especially in response to treatment. 

Frequent use of vibration perception testing occurs during routine screenings of people experiencing peripheral neuropathy. 
The most common clinical diagnostics for peripheral neuropathy are the Semmes-Weinstein monofilament exam and the 128~Hz tuning fork exam~\cite{pop2017diabetic}. Although these exams provide important information on tactile ability, they have limited measurement resolution. Tuning forks often provide inconsistent vibrations due to differences in how the clinician strikes the fork~\cite{rossing1992acoustics}. Monofilaments suffer from variations in force delivered due to variations in clinician application and overuse~\cite{dros2009accuracy}. Also, when used in many clinical settings, both of these tools only measure a binary response (`\textit{yes, can feel}' or `\textit{no, cannot feel}') to a single provided stimulus. 

Given the rise of smartphones with high-quality vibration actuators, there has been a rising interest in conducting mobile haptics experiments~\cite{yoshida2023cognitive}~\cite{blum2019getting}. We conducted preliminary work which showed that smartphone vibrations are more repeatable than tuning fork vibrations and that vibration perception threshold (VPT) could be reliably measured using a custom application employing a staircase algorithm~\cite{adenekan2022feasibility}. Also in 2022, Torres et al.{~\cite{torres2023skin} characterized smartphone vibrations and \revision{found that
} smartphone-based VPTs correlate with monofilament-based pressure thresholds in the index finger. 
Several researchers have also worked toward validating the use of smartphone vibrations for diagnosing diabetic peripheral neuropathy. May et al.~\cite{may2017mobile} showed that vibrations generated from a mobile phone could be used to detect diabetic peripheral neuropathy, and that the most accurate testing location was the first metatarsal head (a bony prominence in the big toe). Jasmin et al.~\cite{jasmin2021validity} found that a vibration-based smartphone application has a moderate to strong correlation to tuning forks in classifying participants as experiencing neuropathy or not experiencing neuropathy, and that the \revision{interrater
} reliability using the smartphone application is high. While these studies are important steps towards developing an improved measurement tool, they suffer from various limitations including confounding factors in the study design, non-autonomous smartphone vibration perception data collection (reliance on an experimenter to physically conduct the exam), and the use of a binary measurement (yes/no) as opposed to a continuous, numerical value.

\begin{figure*}[h]
     \centering
    \includegraphics[width=1.0\textwidth] {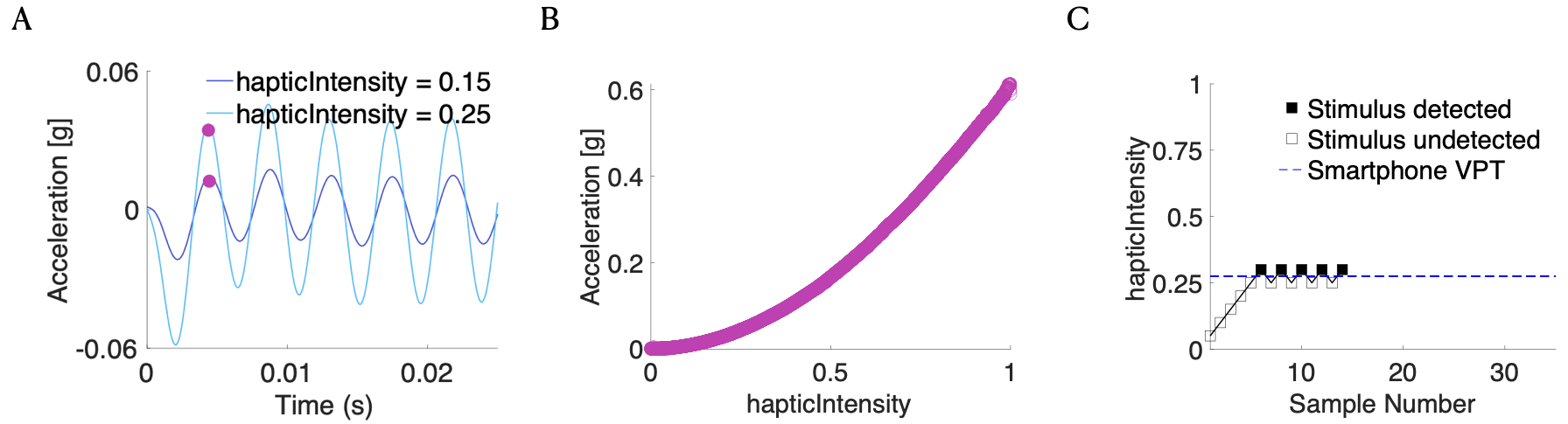}
  \caption{(A) \revision{
  } \revision{
  F}iltered \revision{smartphone} vibration acceleration data for \textit{hapticIntensities} of 0.15 and 0.25 using the setup described in detail in~\cite{adenekan2022feasibility} and~\cite{yoshida2023cognitive}\revision{, with the slight modification that the phone was placed on a pillow, not held in the hand}. (B) Peak accelerations of the waveforms for each \textit{hapticIntensity} (indicated with pink points in (A)). (C) Sample perception data for one trial of the smartphone VPT exam showcasing the staircase method.}
  \label{fig:phone-waveform}
  \vspace{-0.2in}
\end{figure*}

In this work, we build upon our preliminary work by testing the feasibility of an application designed to measure smartphone-based vibration perception thresholds at different sites on the body (hand, wrist, and foot) in a healthy, age-diverse population. We also assess how the smartphone-based vibration perception thresholds correlate to tuning fork vibration perception thresholds and monofilament force perception thresholds at these locations. Our main goal is to confirm that smartphones can be used for vibration perception measurements and capture similar trends as currently used methods. In Section~\ref{sec:methods}, we describe our methods in characterizing, designing, and administering the smartphone-based VPT, performing the tuning fork and monofilament exams, and conducting a user study and accompanying analyses. We then discuss the study results in Section~\ref{sec:results} and key takeaways and future work, such as testing our smartphone application on various patient populations and expanding this application to additional types of smartphones, in Section~\ref{sec:conclusion}.

\section{Methods}
\label{sec:methods}
\subsection{Measurement Methods}
\subsubsection{Smartphone VPT Exam}
We developed an iOS application that controls Apple's Core Haptics parameters (\textit{hapticIntensity} and \textit{hapticSharpness}) and autonomously implements a staircase algorithm (reversals = 8) to measure VPT. \revision{We chose to implement a one-up/one-down staircase method because it is a simple and fast way to calculate an absolute threshold. The one-up/one-down staircase method targets a performance level of 50\%~\cite{levitt1971transformed}~\cite{leek2001adaptive}, which is the standard performance level for absolute thresholds~\cite{levitt1971transformed}.  We chose to implement a one-up/one-down staircase instead of alternative one-up/two-down or one-up/three-down staircase methods that target higher performance levels because these alternatives require more trials and therefore more time to calculate a threshold; we chose to prioritize time as our eventual goal is to use this tool in a clinical setting where time is limited.} Prior work by Yoshida and Kiernan, et al.~\cite{yoshida2023cognitive} characterized the acceleration outputs at various locations on the phone; their results indicate that the output accelerations occur at a constant frequency of 230~Hz and that there is a positive, nonlinear relationship between the amplitude of the output acceleration and commanded \textit{hapticIntensity} value when \textit{hapticSharpness} is held constant at 1 and \textit{hapticIntensity} is varied between 0.1 and 0.3. We further characterized this nonlinear relationship when \textit{hapticSharpness} is held at a constant value of 1 and \textit{hapticIntensity} is varied between 0 and 1 (Fig.~\ref{fig:phone-waveform}). 

Despite this minor nonlinearity, we designed our staircase algorithm to output a continuous vibration for 0.1 seconds with the \textit{hapticSharpness} set at 1.0 and the \textit{hapticIntensity} varying with each step. This results in repeatable vibrations of varying amplitude with a constant frequency of 230 Hz, to which users can quickly and easily respond (Fig.~\ref{fig:phone-waveform}). We chose to use \textit{hapticSharpness} = 1 (230 Hz) since the Apple iPhone XS Max (our phone model) uses \revision{a} linear resonant actuator\revision{
} that \revision{is
} tuned to operate at \textit{hapticSharpness} = 1 (230 Hz). \revision{In addition to 230 Hz being the resonant frequency for this phone model's linear resonant actuator, this frequency is also \revisiontwo{near the maximum sensitivity frequency 
of the Pacinian corpuscles, the mechanoreceptors 
largely} responsible for our ability to feel \revisiontwo{high-frequency} vibrations ~\cite{verrillo1992vibration}. 
}

The vibration amplitude starts small (\textit{hapticIntensity} = 0.05) and increases with a step size of 0.05 until the vibration is detected by the user. At this point, the user says `\textit{yes}' to indicate detection, the spoken `\textit{yes}' is interpreted by the app using ``Speech'' (Apple's voice recognition framework), and a reversal is recorded in the app. The vibration's amplitude then decreases with a step size of 0.05 until the user can no longer detect the vibration (does not provide a spoken `\textit{yes}' response). Then, another reversal is recorded, the vibration's amplitude increases again, and the staircasing of the \textit{hapticIntensity} values continues until eight reversals are complete. Once complete, a CSV file containing the trial data is exported and stored in Google Firebase. The vibration perception threshold is calculated by averaging the \textit{hapticIntensity} values at the eight reversal points (where the reversal point is the average of the value of the response that triggered a reversal and the value of the response prior to the reversal) as shown in Fig.~\ref{fig:phone-waveform}. Time intervals between vibrations for each trial were randomly selected to reduce bias (ranging from 3-6~s), and responses had to occur within 2.5 seconds of the vibration in order to be recorded as a true positive response (as this is the upper end of haptic response times reported in literature~\cite{peon_reactiontime, oldmen_reactiontime}). \revision{The smartphone VPT ranges from \textit{hapticIntensity} of 0.05 (smallest vibration detected) to 1 (largest possible vibration output via Apple’s framework). As such, lower scores indicate better perception. If the largest vibrations output by the phone (vibrations generated by setting \textit{hapticIntensity} to 1) were not felt three times in a row, the threshold for that trial is recorded as NaN because their threshold was outside the range of the smartphone-generated vibration stimuli.}

We measure VPT at six locations of interest: the index finger pad, the back of the index finger, the pinky finger pad, the \revisiontwo{
dorsal} wrist, the \revisiontwo{
volar} wrist, and the big toe pad (Fig.~\ref{fig:phone-positions}). These locations were chosen both for clinical relevance and ease of smartphone placement~\cite{yang2018simple}. \revision{In Fig.~\ref{fig:phone-positions}B we also include the filtered vibration accelerations for the six configurations shown in Fig.~\ref{fig:phone-positions}A. For these measurements \textit{hapticIntensity} was set to 0.25 and \textit{hapticSharpness} was set to 1. We measure these waveforms using the setup described in detail in~\cite{adenekan2022feasibility} and~\cite{yoshida2023cognitive}, with the modification that the phone was placed on a pillow while the body part was placed on the phone in the six different configurations.} For each participant, smartphone-based VPT measurements are collected five times at each location, so that we can report the participant's average smartphone-based VPT at each location. 

All smartphone-based measurements are collected on an Apple iPhone XS Max. Participants sit in a chair during the entire exam. For the finger and wrist locations, the phone is placed on a pillow that is placed on a desk. For the foot location, the phone is placed on a pillow, that is placed on the floor. Participants wore headphones playing their preferred music to prevent the use of auditory cues to identify vibrations.  All but two participants listened to a Disney Hits playlist. Because the smartphone data collection component of the experiment took around an hour for most participants, we did not use white noise, which was found to make participants drowsy during piloting. \\

\begin{figure}[]
     \centering
     \includegraphics[width=1.0\columnwidth]{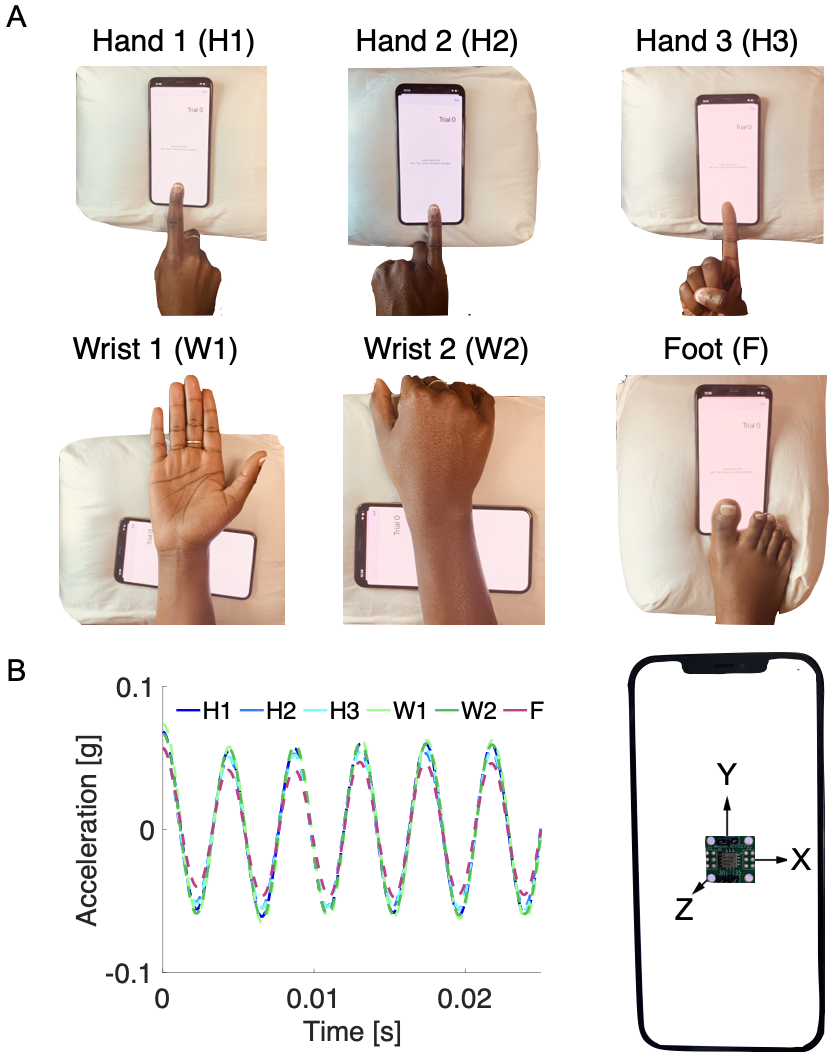}
  \caption{\revision{(A)} The six body locations were tested with all three measurement methods. The smartphone was placed on a pillow and participants were asked to contact the phone as shown in this figure. During the monofilament and tuning fork tests, participants sat in a chair, and a pillow resting on a desk or coffee table allowed participants to comfortably support their limbs. \revision{(B) The filtered vibration waveforms for the six configurations shown in (A). For these measurements \textit{hapticIntensity} was set to 0.25 and \textit{hapticSharpness} was set to 1.} }
  \label{fig:phone-positions}
  \vspace{-0.2in}
\end{figure}

\subsubsection{Tuning Fork Exam}
We use a 128 Hz clinical tuning fork (CynaMed) to test the same six body parts as the smartphone (Fig.~\ref{fig:phone-positions}). Participants sit in a chair during the entire exam. For the finger and wrist locations, the hand/wrist is placed on a pillow that is placed on a desk. For the foot location, the foot is placed on a pillow, that is placed on a coffee table. The experimenter strikes the tines of the tuning fork on her knee and then places the base of the tuning fork on the body of the participant. Prior to collection, participants are touched with a vibrating tuning fork so they could get a sense of what it feels like and so that they understand that the sensation is not painful. Participants wear a blindfold so that they can not see when the tuning fork makes contact with their body, but do not wear headphones as the sound of the tuning fork vibrations are not easily discernible. Participants are instructed to say ``start" when they start feeling a vibration and ``stop" when they no longer feel any vibrations. A digital watch displaying seconds is used to measure the amount of time that the participant feels the vibration. The experimenter notes the second when the participant says ``start'' and also the second when the participant says ``stop'' and then records the difference in seconds, mimicking methods used in clinical settings. \revision{As such, higher tuning fork perception times indicate better perception.} At each body part, tuning fork vibration perception time is measured five times. The times are then averaged, so that each participant's average perception time at each of the body parts is reported.\\

\subsubsection{Monofilament Exam}
We use a 20-piece Semmes-Weinstein monofilament set (Touch Test Sensory Evaluators, North Coast Medical, Inc.) to assess light force perception. Monofilaments range from 0.008 grams-force to 300 grams-force. To assess light force perception at each of the six body parts (Fig.~\ref{fig:phone-positions}), we begin with the monofilament deemed normal for that location (0.07 grams-force for hands and dorsal feet, 0.4 grams-force for the plantar feet)~\cite{TouchTes36:online}. Participants sit in a chair during the entire exam. For the finger and wrist locations, the hand/wrist is placed on a pillow that is placed on a desk. For the foot location, the foot is placed on a pillow, that is placed on a coffee table. Prior to collection, participants were touched with a sample monofilament so they could get a sense of what it feels like, and so that they understand that the stimulus is not painful. Participants wear a blindfold so that they can not see when the monofilament makes contact with their body, but do not wear headphones as the stimuli are inaudible. We then follow the protocol provided with the monofilament kit, mimicking how the procedure would be performed by clinicians~\cite{TouchTes36:online}. The participants are instructed to say ``yes" anytime they feel the monofilament touching them. \revision{If the participant does not feel the monofilament, we increase the monofilament evaluator size, otherwise we decrease the monofilament evaluator size.
} Once the participant does not feel a monofilament size (after three touches with the 0.008 grams-force to 1.000 grams-force monofilaments or one touch with the 1.400 grams-force to 300 grams-force monofilaments), we record the minimum evaluator size that they could feel as their monofilament threshold for that location. If the participant does not feel the starting monofilament, we increase the evaluator size until the subject can feel it and then record that evaluator size as their monofilament threshold for that location. \revision{As such, lower monofilament thresholds indicate better perception.} 

\begin{table}[htbp]
\caption{Participant Demographics, \revision{N = \revisiontwo{28 
} 
\revisiontwo{\tablefootnote{\revisiontwo{Although we tested 36 participants, we only present data from 28 participants. Eight participants are excluded from the table; their results were not used in the analyses as explained in Section~\ref{subsec:stats}.}}}
}}
\begin{center}
\begin{tabular}{c|c|c}
\textbf{Sex assigned at birth}& Male&  \revisiontwo{12 
}  \\
& Female&  \revisiontwo{16 
}  \\
\hline
\textbf{Age}& Older Adults (over 50)&  \revisiontwo{15 \revisiontwo{
} 
}  \\
& Younger Adults (18-50)&  \revisiontwo{13 
}  \\
& \revision{Older Mean (Range)}&  \revision{6
\revisiontwo{7}
(\revisiontwo{55
},80)} \\
& \revision{Younger Mean (Range)}&  \revision{2
\revisiontwo{8}
(22,37)}  \\
\hline
\textbf{Race/Ethnicity}& American Indian / Alaska Native&  1  \\
& Hispanic / Latino&  \revision{6 
} \\
& Black / of African Descent&  \revisiontwo{2 
} \\
& Asian&  \revisiontwo{6 
}  \\
& White&  \revisiontwo{18 
}  \\
& Other (Brazilian)&  1  \\
\hline
\revision{\textbf{Handedness}}& \revision{Right}&  \revisiontwo{26 
} \\
& \revision{Left}&  \revisiontwo{1 
} \\
& \revision{Ambidextrous}&  \revisiontwo{1 
} \\
\hline
\revision{\textbf{Footedness}}& \revision{Right}&  \revisiontwo{25 
} \\
& \revision{Left}&  \revisiontwo{1 
} \\
& \revision{Ambipedal}&  \revision{1} \\
& \revision{Unknown}&  \revisiontwo{1 
} \\
\end{tabular}
\label{demographics}
 \vspace{-0.2in}
\end{center}
\end{table}

\subsection{User Study Design}
\subsubsection{Participants}
\revisiontwo{
Thirty-six} adult participants with no known history of diabetes or other disorders linked to peripheral neuropathy completed this study. Participant demographics are displayed in Table~\ref{demographics}. This study was approved by the Stanford University Institutional Review Board under Protocol 22514, and written consent was provided by all participants. Prior to completing the study, participants completed a pre-survey that inquired about demographic information as well as hobbies or injuries that may impact touch sensitivity at the hands or feet. \revisiontwo{Hand and foot dominance (defined as the foot one would use to kick a ball) were also inquired about via the survey.} \\

\begin{figure*}[!ht]
     \centering
         \includegraphics[width=1.0\textwidth]{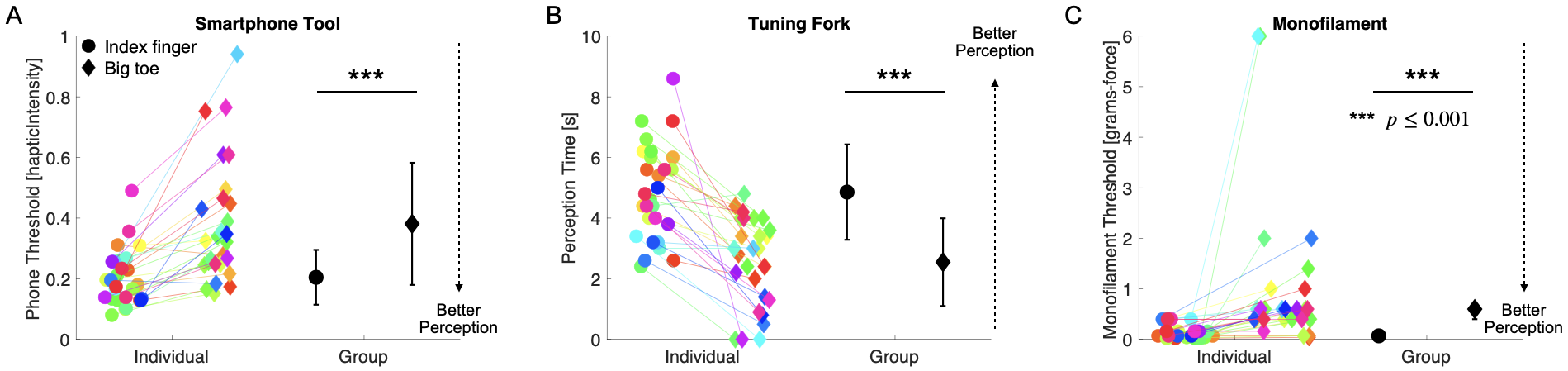}
  \caption{Perception thresholds at the index finger and the big toe for measurement method. Individual thresholds (left) as well as group mean and standard deviation (right) are shown for the smartphone (A) and tuning fork (B). Individual thresholds (left) as well as group median and 25th and 75th quantiles (right) are shown for the monofilaments (C). All three modalities detect statistically significantly \revisiontwo{
  poorer perceptual resolution} at the big toe than at the index finger.}
  \label{fig:bodyparts}
\end{figure*}

\subsubsection{Procedure}
Participants completed a two-day protocol with a one-hour session each day. The same time block was used for each day. Monofilament and tuning fork perception data were collected in the same session on one day and smartphone perception data was collected in the session on the other day. The ordering of the sessions was randomized, and the ordering of the monofilament and the tuning fork data collection within the session was also randomized. For each given modality, the ordering of the body parts tested was randomized. All perception measurements were collected on the participant's dominant side body parts.

\subsection{Statistical Analyses}
\label{subsec:stats}
Perception data obtained from the smartphone, tuning fork, and monofilaments are presented as both individual and group-level data. Smartphone, tuning fork, and monofilament thresholds for \revision{
}\revision{6 participants} were removed \revision{from further analysis} due to the presence of too many false positives during the monofilament \revision{or tuning fork} exam (saying they felt a touch \revision{or vibration} from the filament \revision{or tuning fork} when it was not touching them). \revision{Smartphone, tuning fork, and monofilament thresholds for two participants were removed from further analysis as a different tuning fork was used during their trials.} \revision{This resulted in 13 younger adults and 15 older adults with usable data.} Group level, smartphone-based data and tuning fork-based data are presented as means and standard deviations as smartphone and tuning fork thresholds are both continuous data. Group level, monofilament-based data are presented as medians and quantiles (25th and 75th) as monofilament thresholds are discrete. \revision{
}

Differences in \revisiontwo{
thresholds measured} at the index finger and big toe (H1 and F from Fig.~\ref{fig:phone-positions}) are \revisiontwo{
analyzed} using paired single-sided t-tests with Bonferroni correction for the smartphone and tuning fork and a single-sided Wilcoxon signed rank test (the nonparametric equivalent) for the monofilament.  

Differences in \revisiontwo{
thresholds measured} between younger adults and older adults at the index finger are \revisiontwo{
analyzed} using unpaired single-sided t-tests with Bonferroni correction for the smartphone and tuning fork and a single-sided Wilcoxon rank sum test with Bonferroni correction (the nonparametric equivalent) for the monofilament. Differences in \revisiontwo{
thresholds measured} between younger adults and older adults at big toe are \revisiontwo{
analyzed} in the same manner.

\revision{The corresponding effect sizes are reported for each statistical test. Namely, Cohen's d effect sizes (d) are reported for the t-tests, and Wilcoxon effect sizes (r) are reported for the Wilcoxon tests.}

Correlations between the three modalities are performed using Pearson's correlation coefficient for smartphone and tuning fork and Spearman's rank correlation coefficient for smartphone and monofilament as well as monofilament and tuning fork to account for the discrete monofilament thresholds. While Pearson's correlations result in a line of best fit, Spearman's correlations do not result in a line of best fit because it is used to describe \revision{monotonic, not necessarily linear 
}relationships.  \revision{
}

All statistical analyses except quantile \revision{and Wilcoxon effect size} calculations for the monofilament data are conducted using MATLAB 2022b. The quantile \revision{and effect size} calculations for the monofilament data are performed using R (Version 4.0.3)\revision{. 
This} allows for quantiles to be calculated for discrete and non-normal data that are unsuitable for linear interpolation. We used R's type 1 quantile calculation algorithm.

\begin{figure*}[!ht]
     \centering
         \includegraphics[width=1.0\textwidth]{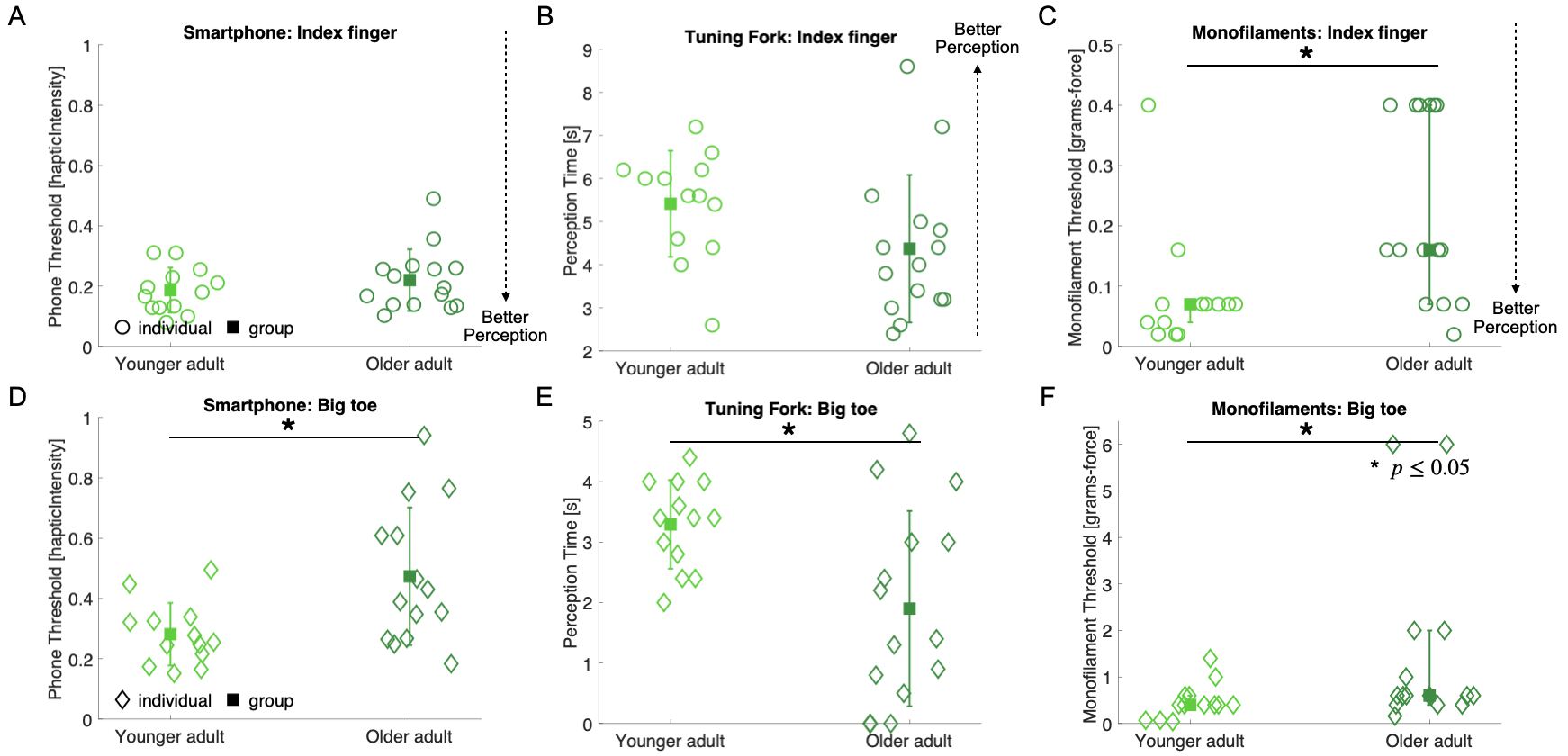}
\caption{Perception thresholds in younger adults (left) and older adults (right) at the index finger (top) and big toe (bottom). Individual thresholds as well as group mean and standard deviation are shown for the smartphone (A) and tuning fork (B). Individual thresholds as well as median and 25th and 75th quantiles are shown for the monofilaments (C).}
\label{fig:age}
\vspace{-0.2in}
\end{figure*}

\section{Results and Discussion}
\label{sec:results}
\subsection{Discrimination Between Hands and Feet}
\label{subsec:disLoc}
To investigate the resolution limits of our smartphone-based tool, we first sought to determine whether our tool could detect known trends in vibration perception of different body locations in healthy humans. As shown in Fig.~\ref{fig:bodyparts}A, our participants had a smartphone \revision{threshold
} value of \revision{$0.20\pm{0.09}$
}~\textit{hapticIntensity}\revision{
} ($mean\pm{std}$) at the index finger and \revision{$0.38\pm{0.20} $
}~\textit{hapticIntensity}\revision{
} at the big toe. From our statistical analysis, we found that our smartphone-based tool detected a statistically significant VPT difference between the index finger and big toe (\revision{$p=\num{1.28E-5}$
}). \revision{We also found that our smartphone-based tool detected a large effect size between the index finger and the big toe ($d = -1.09$).}

For clinical comparison, we conducted these same analyses on the tuning fork perception data. Tuning fork perception values were \revision{$4.86\pm{1.57}$
}~s at the index finger and \revision{$2.55\pm{1.45}$~s 
}at the big toe (Fig.~\ref{fig:bodyparts}B). Similar to the smartphone, there was a statistically significant difference between index finger and big toe (\revision{$p=\num{3.28E-7}$
})
in the tuning fork threshold measurements. \revision{We also found that the tuning fork detected a large effect size between the index finger and the big toe ($d = 1.49$).}

\begin{figure*}[!bt]
     \centering
         \includegraphics[width=0.95\textwidth]{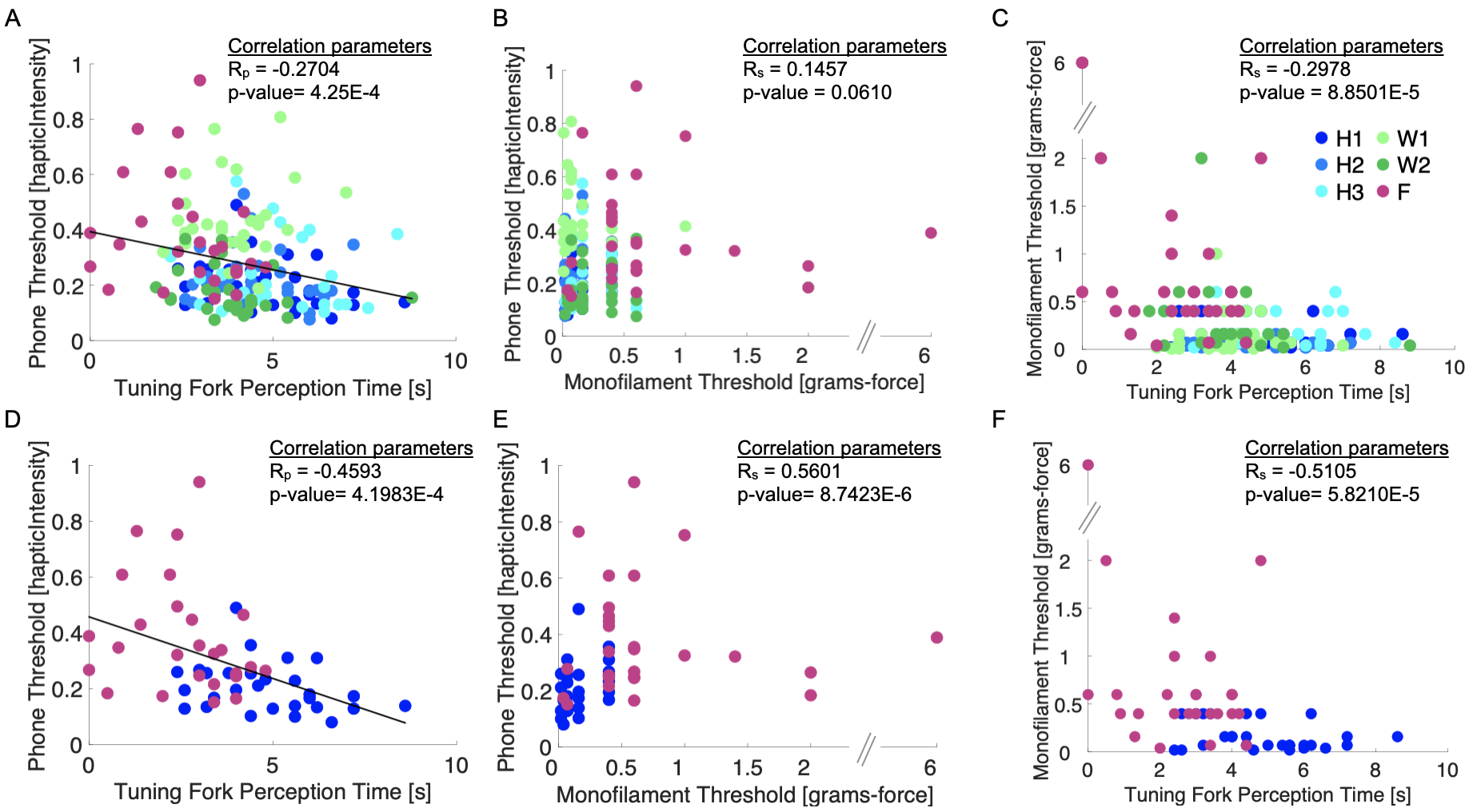}
  \caption{Correlations between each of the measurement methods. Pearson's correlation coefficient is shown for the smartphone and tuning fork data (A). Spearman's correlation coefficient is shown for correlations between the smartphone and monofilaments data (B) and between the monofilaments and tuning fork data (C). As a reminder, (B) and (C) do not have best fit lines because Spearman's correlation is used to describe \revision{monotonic, not necessarily linear} 
  relationships. \revision{(D), (E), and (F) are the corresponding figures for (A), (B), and (C), when only the index finger (a convenient location for perception testing) and big toe (a clinically relevant location) are included.}}
  \label{fig:correlations}
\end{figure*}

We conducted similar analyses using nonparametric equivalents on the monofilament perception data. The monofilament force perception values were as follows: \revision{0.07
[0.07, 0.16]~grams-force 
}($median$ [$25th$, $75th$ $quantiles$]) at the index finger and \revision{0.6 [0.4, 0.6]~grams-force 
}at the big toe (Fig.~\ref{fig:bodyparts}C). We again found that there was a statistically significant difference between the monofilament threshold measurements at the index finger and at the big toe (\revision{$p=\num{2.85E-5}$
}). \revision{We also found that the monofilaments detected a large effect size between the index finger and the big toe ($r = -0.854$).}

In short, all three modalities yielded data that align with previous research findings that hands are more sensitive to vibrations and force than feet~\cite{morioka2008vibrotactile}. The ability to replicate known trends further strengthens the hypothesis that smartphones may be able to provide high enough resolution data to classify vibrotactile perception at an even finer scale than just neuropathic or non-neuropathic.

\begin{figure*}[!ht]
     \centering
         \includegraphics[width=0.8\textwidth]{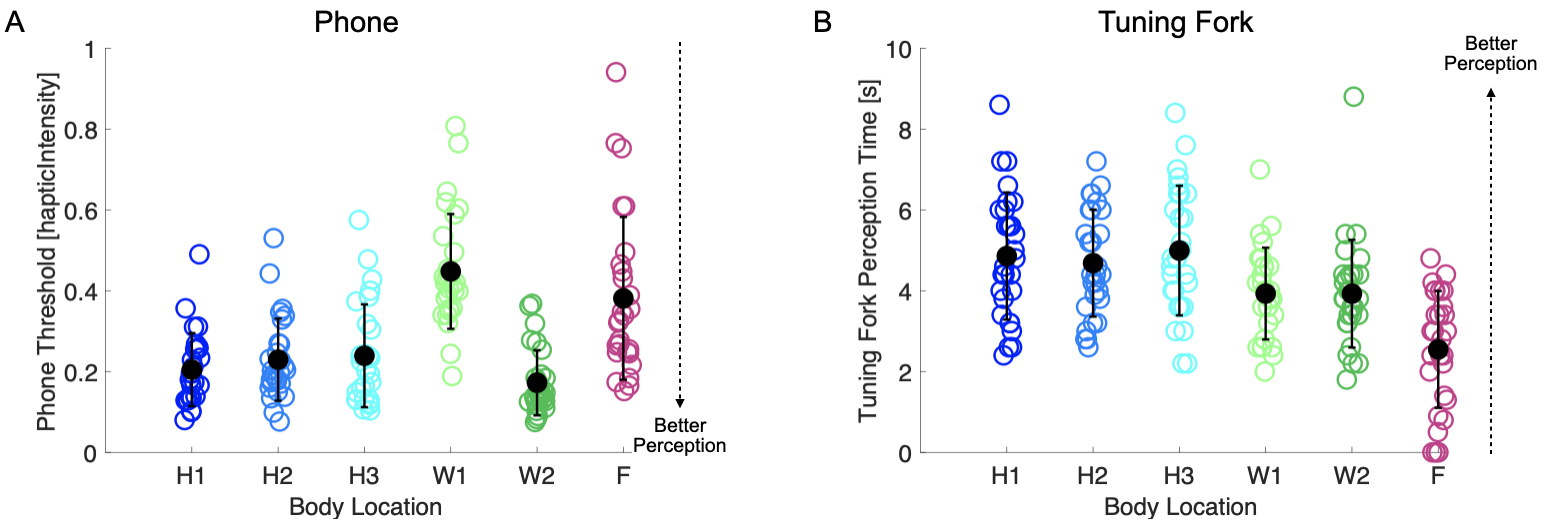}
  \caption{\revision{Threshold vs. body location for the smartphone (A) and tuning fork (B). Colored circle outlines indicate individual threshold values, black-filled circles represent the group mean, and the bars indicate the standard deviation. The trend across body parts for the smartphone is inverted from that of the tuning fork, as expected, with the exception of the W2 location. }}
  \label{fig:scatter-allPos}
  
\vspace{-0.2cm}
\end{figure*}

\subsection{Discrimination Between Younger and Older Adults}
We also investigated the smartphone's ability to measure known effects of age on vibrotactile perception (Fig.~\ref{fig:age}A). From our statistical analyses, we found that the smartphone could not discriminate between younger (\revision{$0.19 \pm {0.07}$~\textit{hapticIntensity}
}) and older ($0.22\pm{0.10}$~\textit{hapticIntensity}\revision{
}) adults at the index finger location (\revision{$p=0.52$
}) but could discriminate between younger (\revision{$0.28\pm{0.10}$~\textit{hapticIntensity}
}) and older \revision{($0.47\pm{0.23}$~\textit{hapticIntensity}}\revision{
}) adults at the big toe ($p=0.02$).
\revision{We also found that the smartphone detected a small effect size between the index fingers of the younger and older adults ($d = -0.35$). There was a large effect size between the big toes of the younger and older adults ($d = -1.04$). }

We ran the same statistical tests on the tuning fork exam data for clinical comparison. As shown in Fig.~\ref{fig:age}B, we found that the tuning fork exam could \revision{not} discriminate between younger ($5.42\pm{1.23}$~s) and older ($4.37\pm{1.71}$~s\revision{
}) adults at the index finger location (\revision{$p=0.12$
}). However, the tuning fork exam could\revision{
} discriminate between younger ($3.29\pm{0.73}$~s) and older (\revision{$1.90\pm{1.61}$~s
}) adults at the big toe location (\revision{$p=0.01$
}). \revision{We also found that the tuning fork detected a moderate effect size between the index fingers of the younger and older adults ($d = 0.67$). There was a large effect size between the big toes of the younger and older adults ($d = 1.05$). }

We conducted similar analyses using the nonparametric equivalents on the monofilament perception data (Fig.~\ref{fig:age}C). The monofilament exam could \revision{
}distinguish between \revision{both} younger (\revision{0.07 [0.04, 07]~grams-force
}) and older (\revision{0.16 [0.07, 0.40]~grams-force
}) adults at the index finger (\revision{$p=0.01$
}), \revision{ and 
} could identify a significant difference at the big toe between younger (\revision{0.40 [0.40, 0.60]~grams-force
}) and older (\revision{0.60 [0.40, 2.00]~grams-force{
}}) adults (\revision{$p=0.04$
}). \revision{There was a moderate effect size between the index fingers of the younger and older adults ($r = -0.54$). There was a moderate effect size between the big toes of the younger and older adults ($r = -0.42$). }

To summarize, the smartphone was able to replicate the known trend of older adults having worse vibrotactile perception than younger adults in the feet~\cite{wells2003regional}. The \revision{tuning fork and }monofilament exam\revision{s} \revision{were 
}also able to replicate this same known trend. \revision{
}

\subsection{Correlation Between Smartphone VPT and Clinical Measurements}

To better understand how our smartphone perception values correlate to clinical standards, we calculated correlation coefficients (Fig.~\ref{fig:correlations}). As shown in Fig.~\ref{fig:correlations}A, we found a statistically significant, but weak, \revision{negative
} correlation between the smartphone-based VPT and the tuning fork VPT (\revision{$R_p=-0.270$, $p=\num{4.25E-4}$
}) (Fig.~\ref{fig:correlations}). We also found a \revision{non-}statistically significant, but very weak, positive correlation (\revision{$R_s=0.146$, $p=0.061$
}) between the smartphone-based \revision{VPT 
} and the monofilament \revision{force threshold 
}  (Fig.~\ref{fig:correlations}B). Finally, we found a statistically significant, but \revision{
} weak, \revision{negative 
} correlation between the monofilament \revision{force threshold
} and the tuning fork \revision{VPT 
}(\revision{$R_s=-0.298$, $p=\num{8.85E-5}$
}) as shown in Fig.~\ref{fig:correlations}C. 

The correlation\revisiontwo{
} between the \revisiontwo{
}smartphone and tuning fork is negative as expected. The correlation between the smartphone and monofilaments is positive as expected. 
}Both the smartphone and tuning fork measure vibrotactile ability, so we expect those to be more closely correlated than the monofilaments (which measure force, not vibration perception). However, the correlations are quite weak. One possible reason for this result is that the tuning fork and monofilament perception values are subject to inconsistencies, such as variation in how hard the experimenter struck the tuning fork and how hard the experimenter pressed the monofilaments.
\revision{Another possible reason is that we chose locations that did not yield much spread in the VPT data. To better assess this idea, we calculated correlations for the smartphone VPT vs. tuning fork perception time when one focuses solely on the index finger and foot locations. The index finger is a convenient and common
vibration perception testing location and the foot is an important testing location for clinical applications. We noticed that when one isolates the index finger and foot location, there is a much stronger correlation between the phone and the clinical tools. As shown in Fig.~\ref{fig:correlations}D, we found a statistically significant, and moderate, negative correlation between the smartphone-based VPT and the tuning fork VPT ($R_p=-0.459$, $p=\num{4.20E-4}$). As shown in Fig.~\ref{fig:correlations}E, we found a statistically significant, and moderate, positive correlation between the smartphone-based VPT and the monofilament threshold ($R_s=0.560$, $p=\num{8.74E-6}$). As shown in Fig.~\ref{fig:correlations}F, we found a statistically significant, and moderate, negative correlation between the monofilament threshold and the tuning fork perception time ($R_s=-0.511$, $p=\num{5.82E-5}$). }

\revision{We also visualized how vibration perception changes across different body parts with the phone and the tuning fork in Fig.~\ref{fig:scatter-allPos}. As one would expect, the trend across body parts for the smartphone is inverted from the trend across body parts for the tuning fork. The one exception is the W2 location, and we believe this may be due to a confounding factor of the experimental setup. Later in the data collection, we realized that many participants could feel W2 vibrations in their fingers as well as through the wrist because of how vibrations from the phone are propagated through the pillow to the fingers in the W2 configuration (Fig.~\ref{fig:phone-positions}). This would have caused the W2 smartphone-based VPTs to be more similar to the hand smartphone-based VPTs.}

\revision{
}





\section{Conclusion}
\label{sec:conclusion}
Using Apple’s Core Haptics Framework along with standard psychophysical methods, we developed a smartphone-based vibration perception threshold test that can be used to identify known trends in vibrotactile ability. Fingers are known to be more touch-sensitive than feet~\cite{morioka2008vibrotactile}. Additionally, younger adults are known to have more touch-sensitive feet than older adults~\cite{wells2003regional}. \revision{
O}ur smartphone successfully identified these trends. 

We also found that our smartphone-based vibration perception measures were weakly correlated to clinical tuning fork-based vibration perception measures across all \revision{measured body locations
} and \revision{moderately
} correlated to clinical tuning fork-based vibration perception measures \revision{when we isolate the index finger (a convenient testing site) and foot (a clinically relevant site)
}. \revision{
}

\revision{
}

\revision{While we observe a moderate correlation in the thresholds obtained between our smartphone-based tool and the clinical tools (tuning forks and monofilaments), there are a few reasons we don't expect a very strong correlation.  One reason is that the tools use different stimulus directions (staircase for smartphone, ascending series for standard clinical monofilament exam). Perception thresholds are known to vary with stimulus series direction. And while we aimed to obtain a more accurate measure of perception (using staircase method with the phone), we aimed to compare to a standard clinical monofilament exam (which uses an ascending series). Another reason could be that monofilaments measure pressure perception which is different from our smartphone-based tool which measures vibration perception. Differences between the phone and tuning fork may be attributed to the fact that the tuning fork is difficult to hit reliably. The vibrations vary with how much force the fork was stuck with and also how much time the experimenter took to place the fork on a given body site.}
The aim of this initial work was to confirm the feasibility of the iPhone XS Max and app platform to accurately measure vibration perception thresholds. \revision{To more rigorously test the clinical relevance of our tool, in future research we will compare our smartphone-based vibration perception threshold measurements to measurements of sensory function as measured by nerve conduction studies and Rydel Seiffer tuning forks which are more specialized clinical tools that provide higher fidelity measurements than the clinical tuning fork and monofilament exam.} \revision{
W}e will \revision{also} conduct \revision{larger} studies comparing \revision{healthy controls and participants with 
}different levels of risk of developing \revision{peripheral} neuropathy. This will enable us to assess whether our platform can be used to measure progression and regression of nerve damage. If successful, our platform could be used to identify those at risk of developing irreversible nerve damage, and could motivate at-risk individuals to adhere to treatment and management plans. Given the ubiquitous nature of smartphones, the tool could be used both in and outside of clinics to increase access to reliable sensory diagnostic tests. However, different types of smartphones contain varying hardware and control variables. Thus, in order to make this vision a reality, future work must be done to characterize additional smartphones and expand the make and models of smartphones that can be used within our platform.
 

\section*{\revision{Acknowledgments}}
\revision{The authors thank \revisiontwo{the participants for their time and efforts in providing us with a rich dataset, J. Mucini, M.S., OTR/L for her clinical advice regarding the monofilament and tuning fork protocols,} Bryce N. Huerta, M.S. and Alexis J. Lowber, M.S. for their contributions to the early iPhone application development, \revisiontwo{and} Adeyinka E. Adenekan, Ph.D. for insightful discussions regarding the data visualization.}

\bibliographystyle{IEEEtran}
\bibliography{mybibliography.bib}

\vspace{12pt}

\end{document}